# Remote lensless focusing of a light beam

Nikolai I. Petrov

*Lenina street, 19-39, Istra, Moscow region, Russia 143500*

*Corresponding author:* petrovni@mail.ru

Remote focusing of light in a graded-index medium via mode interference is demonstrated using exact analytical solutions of the wave equation. Strong focusing of light occurs at extremely long distances and it revivals periodically with distance due to mode interference. High efficiency transfer of a strongly focused subwavelength spot through optical waveguide over large distances takes place with a period of revival. Super-oscillatory hot-spots with the sizes which are beyond the conventional Abbe diffraction limit can be observed at large distances from the source. This can provide the possibility to detect optical super-resolution information in the far-field without any evanescent waves. Far-field super-resolution imaging capabilities of a graded-index waveguide are also analyzed.

Focusing of light is essential to many research and application areas including microscopy, lithography, optical information storage, optical trapping, etc. Various instruments for focusing and imaging are widely used in biology, medicine, chemistry and material engineering. Although numerous techniques exist to focus of a light, the refractive lenses are most commonly used. Resolution limitations of conventional optics are due to the wave nature of light and it cannot be focused beyond the Abbe-Rayleigh limit [1]. Higher resolution imaging requires the systems with high numerical apertures, which means that the focusing spot will be not far from the lens. If the diffraction limited spot $\delta x$ is of the order of wavelength $\lambda$ then the focusing distance $z_f = 0.82 D\delta x / \lambda$ is of the order of the incident beam diameter *D*. However, there are instances where focusing with lenses is impossible, for example, in distant places or in case of the absence of the sufficient space for a lens arrangement. Besides, often the super-resolution information is necessary in the far-field. Near field approaches [2, 3] can create sub wavelength spots, but they are limited in many applications as the focus spot is in the near field of the lens.

Recently the methods to generate sub wavelength focal spots in the far-field region were considered. In [4] a technique based on the high effective refractive index exhibited by a plasmon-polariton propagation along a metal-dielectric interface is proposed. An approach to transfer an image to far-field region based on a layered uniaxial metamaterial or crystal is demonstrated in [5]. In [6, 7] the focusing of light into subwavelength spot using an optical eigenmode approach implemented with a spatial light modulator and an optical microscope is reported. The phenomenon of superoscillation [8, 9] for super-resolution focusing and imaging without evanescent waves holds much promise. Experimental observations of subdiffraction limited hot spots at a distance of 10 μm from the nanohole array using superoscillations have been reported [10, 11]. However, the focusing distances considered are still small to spatially separate the light source and focusing plane. Besides, in super-oscillation the focusing efficiency significantly decreases as the spot size is reduced [12].

In this paper, the remote focusing of a light beam in a graded-index medium (optical waveguide) is demonstrated using exact analytical solutions of the wave equation and computer simulations. For paraxial beams, the different modes or rays turn to be periodically in phase during their propagation in graded-index optical waveguides, and periodical focusing and defocusing can be observed. However, nonparaxial effects destroy this periodical focusing. Surprisingly, a strong focusing occurs at extremely long distances due to the revival effect caused by the interference between propagating modes. High efficiency transfer of a strongly focused subwavelength spot through optical waveguide over large distances with a period of revival is shown. Super-oscillatory hot-spots with the sizes smaller than $\lambda/(2NA)$ which are beyond the conventional Abbe diffraction limit can be observed at extremely large distances from the source. Far-field super-resolution imaging capabilities of a graded-index waveguide are also analyzed.

The propagation of radiation in one dimensional media (e.g. planar waveguide) is described by the Helmholtz equation for the monohromatic component of electric field $\vec{E}(x,z)$ following from Maxwell equations:

$$\frac{\partial^2 \vec{E}}{\partial x^2} + \frac{\partial^2 \vec{E}}{\partial z^2} + k^2 n^2(x,z)\vec{E} = 0\,, \qquad (1)$$

where $k = 2\pi/\lambda$ is the wavenumber, $n(x,z)$ is the refractive index of medium.

The absence of the medium nonlinearity and loss is assumed below.

In the case of the homogeneous medium in the longitudinal direction z the equation (1) may be reduced to the equivalent Shrodinger equation for the reduced field $\psi(x)$:

$$\hat{H}\psi(x) = \varepsilon\psi(x) \qquad (2)$$

where $\varepsilon$ and $\psi(x)$ are the eigenvalue and eigenfunction of the Hamiltonian

$$\hat{H} = -\frac{1}{2k^2}\frac{\partial^2}{\partial x^2} + \frac{1}{2}\left(n_0^2 - n^2(x)\right) \qquad (3)$$

Evolution of the field $E(x,z)$ is determined by the propagation constant $\beta(\varepsilon)$:

$$E(x,z) = \hat{U}\psi(x),\ \hat{U} = \exp(i\hat{\beta}z)\,, \beta(\varepsilon) = kn_0\left(1 - \frac{2\varepsilon}{n_0^2}\right)^{1/2}.$$

Here the operator of the propagation constant is introduced

$$\hat{\beta} = kn_0 \left(1 - \frac{2\hat{H}}{n_0^2}\right)^{1/2} = kn_0 \left(1 - \frac{\hat{H}}{n_0^2} - \frac{\hat{H}^2}{2n_0^4} - \frac{\hat{H}^3}{2n_0^6} - \cdots\right), \tag{4}$$

the eigenvalues of which determine the propagation constants $\beta(\varepsilon)$.

Note that the eigenvalues spectrum is non-equidistant which leads to the effective nonlinearity.

Thus the solution of the Helmholtz equation (1) in this case may be reduced to the solution of the Heisenberg equation for the operators, the average values of which determine the parameters of the investigated beam, for example, the coordinate and width of the beam.

Consider the propagation of the radiation in a homogeneous medium in the longitudinal direction $z$ with the parabolic distribution of the refractive index in the transverse direction $x$:

$$n^2 = n_0^2 - \omega^2 x^2, \tag{5}$$

where $n_0$ is the refractive index on the axis, $\omega$ is the gradient parameter of the medium.

The wave function of an incident beam is determined by

$$\Psi(x,0) = \left(\frac{2}{\pi}\right)^{1/4} \sqrt{\frac{1}{a_0}} \exp\left(-\frac{(x-\Delta)^2}{a_0^2}\right), \tag{6}$$

where $a_0$ is the width of a beam which is 85% of the full width at half maximum (FWHM), $\Delta$ is the axis displacement.

An incident beam may be expanded into modal functions, so the evolution of a beam in the medium (5) can be represented as

$$\Psi(x,z) = \sum_m c_m \psi_m(x) e^{i\beta_m z}, \tag{7}$$

where $c_0 = \left(\frac{2\sqrt{\omega_- \omega_+}}{\omega_- + \omega_+}\right)^{1/2} \exp\left(-\frac{k\omega_- \omega_+}{2(\omega_- + \omega_+)}\Delta^2\right)$,

$$c_{m+1} = -\frac{v}{u}\left(\frac{m}{m+1}\right)^{1/2} c_{m-1} + \frac{1}{u}\frac{\delta}{\sqrt{m+1}} c_m, \delta = \sqrt{\frac{k\omega_-}{2}}\Delta,$$

$$u = \frac{\omega_- + \omega_+}{2\sqrt{\omega_- \omega_+}}, v = \frac{\omega_- - \omega_+}{2\sqrt{\omega_- \omega_+}}, \quad \omega_+ = \frac{2}{kw_0^2}, \ \omega_- = \frac{2}{ka_0^2},$$

$\psi_m(x)=\left(\frac{k\omega_+}{\pi}\right)^{1/4}\frac{1}{\sqrt{2^m m!}}\exp\left(-\frac{k\omega_+}{2}x^2\right)H_m\left(\sqrt{k\omega_+}x\right)$ are the Gauss-Hermite guided modes of the medium, $w_0$ is the radius of the fundamental waveguide mode, $\beta_m = kn_0\sqrt{1-\frac{2\omega_+}{kn_0^2}\left(m+\frac{1}{2}\right)}$ are the propagation constants. The evolution of the width of the beam $\sigma_x$ is found by calculating the average values

$$\sigma_x^2 = \Delta x_\alpha^2 = \langle x^2\rangle - \langle x\rangle^2, \tag{8}$$

where $\langle x\rangle = \bar{x}(z) = \langle\Psi(x,z)|\hat{x}|\Psi(x,z)\rangle = \langle\Psi(x,0)|\hat{x}(z)|\Psi(x,0)\rangle$.

The average values can be calculated analytically using an operator method [13, 14]. If the axis displacement is zero ($\Delta$ = 0), then we have:

$$\sigma_x^2 = \frac{1}{2k\omega_+}\left\{1+2v^2+\left(\frac{v}{u^2}\right)\sum_{m=0}^{N}\frac{H_{2m}(0)H_{2m+2}(0)}{(2m)!}\left(\frac{v}{2u}\right)^{2m}\cos(\gamma z)\right\}, \tag{9}$$

where $H_m(0)$ are Hermite polynomials. $\gamma = \beta_{2m} - \beta_{2m+2}$.

In the paraxial approximation, expression (9) has the form

$$\sigma_x^2 = \frac{1}{2k\omega_+}\left[1+2v^2-2uv\cos\frac{2\omega_+}{n_0}z\right]. \tag{10}$$

Figures 1(a)-1(d) show the variation with distance of the width of a Gaussian beam with the initial width $a_0 = 20$ μm. The medium (5) with the gradient parameter $\omega$ = 7x10$^{-3}$ μm$^{-1}$ and refractive index $n_0 = 1.5$ is considered. Here and below the beams with wavelength $\lambda$ = 0.63 μm are considered. Numerical aperture $NA = a\omega = n_0\sqrt{2\Delta}$ , where $a$ is the radius of the waveguide and $\Delta \approx (n_0 - n(a))/n_0 = \delta n / n_0 \approx 0.01$, is about 0.2. These parameters are reasonable for conventional graded-index optical fibers. The width of the incident beam oscillates with a period of $L_0 = \pi n_0 / \omega$ and these oscillations relax to the static value determined by the width of an incident beam. Focusing of a beam takes place at a distance of $L_0^f = \pi n_0 /(2\omega)$. It is evident that long-term periodic revivals of the initial width and focusing occur with a period close to $L_{rev} \cong \pi p/(2\eta)$, where $p$ = 1, 2 …, $\eta = \omega^2/(kn_0^3)$. Note that $L_0 << L_{rev}$. Fig. 1c represents the distribution of the field intensity along the axial line of the medium. The field intensity in the focal plane increases by a factor of 13 compared with the intensity in the initial plane.

The wave shape variation with distance is determined by the function $I(x,z) = |\psi(x,z)|^2$, where $\psi(x,z) = \hat{U}\psi(x,0)$. The evolution of beam profiles is shown in Fig. 2. As can be seen from Fig.2, the distributions of the field intensity differ from the initial Gaussian distribution. It is seen that the focused wave shape of a beam revivals periodically at distances $L_{rev}$. The minimum spot size (FWHM) is generally defined as $\delta x \cong \lambda/(2NA)$, where $NA$ is the numerical aperture of a focusing system. For graded-index waveguides with $NA$ = 0.5 the spots with the sizes of the order of λ can be obtained. In the case of higher values of the graded-index parameter we have higher $NA$, so the less focused spots will be received. The higher the difference between the refractive indices on an axis and a boundary of a waveguide the higher the numerical aperture of the waveguide. The focusing efficiency can be defined as the ratio of the power in a spot to the incident beam power. In Fig. 3 the intensity profiles at focusing planes in a waveguide with NA=1 are presented. For the spot size of 340nm the focusing efficiency is equal to 55% which is 65% of the propagating power. Note that in spite of the fact that there are the propagating modes with subwavelength energy localizations (hot-spots) with sizes of $\delta x < \lambda/(2NA)$ which are less than a diffraction limited spot, the minimum focused spot is not less than $\delta x \cong \lambda/(2NA)$. The highest-order propagating mode is described by the Gauss-Hermite function of the order of $m = N_{max}$, where $N_{max} = k\omega a^2/2 = kaNA/2$, and contains super-oscillatory hot-spots with the sizes less than $\lambda/(2NA)$ (Fig. 3c, d). However, they contribute only a small fraction of the total energy, which is only 0.81% of the incident power for $a_0$ = 14 μm and 0.64% for $a_0$ = 45 μm. Super-oscillatory hot-spots can be generated using computer generated holograms (CGHs) designed to excite the higher-order modes of the waveguide. Because of these super-oscillatory hot-spots are the constituents of the propagating modes they can be transferred to a far-field region. Besides, different methods of suppressing or separation the unwanted lower-order modes in a waveguide can be applied. Note, that in a free-space, the hot-spots with the sizes less than the diffraction limited spot $\delta x \cong \lambda / 2$ can be created using a super-oscillatory lens [15].

Consider the propagation of a strongly focused Gaussian beam in a medium (5). Note that only the propagating modes reach the far-field zone. The ratio of the energy guided by propagating modes to the total input energy decreases with the decrease of the incident beam width $a_0$ [16]. For $a_0 \geq \lambda /(2NA)$ all incident power is in the propagating modes and the periodical revivals of the initial Gaussian field intensity distribution occur at extremely long distances. However, in the case of incident beam with a FWHM smaller than $\lambda /(2NA)$ the distribution of the field intensity in focused planes differs significantly from the initial Gaussian distribution, the central spot is surrounded by sidebands. In Fig. 4 the evolution of the field intensity distribution in a transverse plane in a waveguide with *NA* = 0.8 is shown for the incident beam with a FWHM of 141 *nm*. The propagating power is 75% of the incident power, and 54% of propagating power is located in a central part of a focal spot of size of $\delta x \approx \lambda /(2NA) \cong 400nm$. The super-oscillatory hot spots with a FWHM smaller than $\lambda /(2NA)$ can be transferred with the help of propagating modes. In Fig. 4d the intensity profile of the highest-order propagating mode is presented. It is seen that the hot-spots with sizes of $\delta x \cong \lambda /(4NA)$ which are two times less than a diffraction limited spot can be generated in a far-field without the use of the evanescent waves. Although they contain only a small fraction of energy (less than 1.5% of the incident power), their intensity can achieve the value of 10% of the central spot peak intensity. Similar to super-oscillatory lenses [12], where the super-oscillatory hotspots may be realized through a redistribution of intensity from high-frequency to low-frequency modes, here the transformation of energy of a strongly focused beam into the propagating modes of a waveguide allows the subwavelength spot information to be transferred to a far-field region. This indicates that the propagating fields of a strongly focused beam or a light beam passed through the small hole will carry the information about the incident light beam parameters. Note, that the possibility of the far-field characterization of small apertures based on the measurements of the relative propagating mode intensities was experimentally demonstrated in [17].

Let us analyze the imaging capabilities of a graded-index waveguide. An incident beam width can be considered as an object size. For the incident beam width $a_0$ smaller than $\lambda/(2NA)$, the focused spot size does not vary with changing $a_0$. This indicates that the incident beam with the width smaller than $\lambda/(2NA)$ can be considered as a point source, and the intensity distribution on an image (focus) plane will give us the point spread function (PSF). However, the total power in a central spot decreases with the decrease of the size $a_0$ of an object. The survival of the laterally displaced initial state can be examined calculating the autocorrelation function $P(z) = \left|\langle \psi(x,0,\Delta) \,|\, \psi(x,z) \rangle\right|^2$, which measures the correlation of the wave function at the distance $z$ with its initial state at $z = 0$. In Fig. 5 the distributions of the field intensity of three strongly focused beams displaced to each other in the transverse plane are presented. It is seen that as the point object is displaced from the optical axis the corresponding image moves linearly in the transverse direction. This demonstrates the imaging capability of a graded-index waveguide at extremely long distances from the input plane. Image reconstruction takes place periodically with the period of the long-term revivals which in turn consist of periodical short-term repetitions.

Thus, the transfer of a subwavelength focal spot into ultra-long distances in a graded-index medium can be performed with high efficiency. It is shown that the super-resolution information in the far-field can be detected using the propagating modes. This provides super-resolution information without evanescent waves, and therefore without being in the near-field of the object. Although the results obtained correspond to planar optical waveguides, the effect may also manifest itself in two dimensional waveguides (optical fibers). However, the analysis of these effects in this case is more complicated and will include the solution of Maxwell equations taking into account the polarization effects, i.e. the spin-orbit interaction terms in a Hamiltonian [18]. Note that the effects considered will appear in waveguides with loss or gain. It was shown in [13], that the nonparaxial wave beams parameters in a graded-index waveguide with loss or gain are also described by the expression possessing long-term periodicity.

In conclusion, the remote lensless focusing of light in a graded-index medium via mode interference is demonstrated using exact analytical solutions of the wave equation and computer simulations. Strong focusing of light occurs at extremely long distances due to the revival effect caused by mode interference. The transfer of subwavelength focused spots in the graded-index medium is demonstrated. Super-oscillatory hot-spots can be generated in a far-field without the use of the evanescent waves. The far-field super-resolution imaging capabilities of a graded-index waveguide are shown. This can be used to detect the super-resolution information in the far-field. Results obtained may be of great importance in biology and medicine, optical recording and microscopy, and could be exploited in various applications such as novel endoscopes, sensors and imaging systems.

Figure captions:

**Fig.1.** Beam width (a) and axial intensity (c) variations with propagation distance in a waveguide with *NA* = 0.2 and gradient parameter $\omega = 7\times10^{-3}$ μm$^{-1}$, and their higher-resolution plots (b) and (d), respectively. Long-term revivals occur with a period of $L_{rev}$ >> $L_0$, where $L_0 = \pi n_0 / \omega$ is the paraxial beam width oscillation period.

**Fig.2.** Intensity profiles at various distances from the initial plane in a waveguide with *NA* = 0.5 and gradient parameter $\omega = 7\times10^{-3}$ μm$^{-1}$: (a) *z* = 0, (b) *z* = 330.7 μm, (c) *z* = 10 mm, (d) *z* = 1058.8 mm, (e) *z* = 2107.6 mm and (f) *z* = 2128 mm. The initial beam width is $a_0 = 45$ μm, the power in a central spot is 73% of an incident power.

**Fig.3.** Intensity profiles in a waveguide with $\omega$ = 0.1 μm$^{-1}$, *NA* = 1.0 at (a) z = 22 μm and (b) z = 83.15 mm. The initial beam width is $a_0 = 14$ μm. Intensity distribution of the highest-order propagating mode with $N_{max}$ = 50 (c). Higher resolution plot (d) show the super-oscillatory hot-spots with the size of $\delta x \approx 160nm$. The power of the highest-order mode is 0.81% of the incident power.

**Fig.4.** Intensity profiles of a focused beam with the spot size $a_0 = 141nm$ at various distances in a waveguide with $\omega$ = 0.1 μm$^{-1}$, *NA* = 0.8: (a) z = 44.57 μm, (b) z = 200 μm, (c) z = 296.3 mm. Intensity distribution of the highest-order propagating mode with $N_{\max}$ = 32 shows the super-oscillatory hot-spots with the size of $\delta x \approx 160nm$ (d). The power contained in the propagating modes amounts to 75% of the total beam power, and 65-70% of the propagating power is concentrated in a central spot at focused planes.

**Fig.5.** Intensity profiles of axis beam and laterally displaced beams by 1 μm on either side of the optical axis in a waveguide with $\omega$ = 0.2 μm$^{-1}$ and *NA* = 1 at (a) z = 0 – intensity profile of a propagating beam, (b) z = 44.1 μm, (c) z = 2 mm, (d) z = 4.72 mm, (e) z = 10 mm and (f) z =

182.7 mm. The incident beams width is $a_0 = 141nm$, the power in the propagating modes is 80% of the incident power, and 59% of incident power which is 74% of the propagating power in a central spot.

Fig. 1

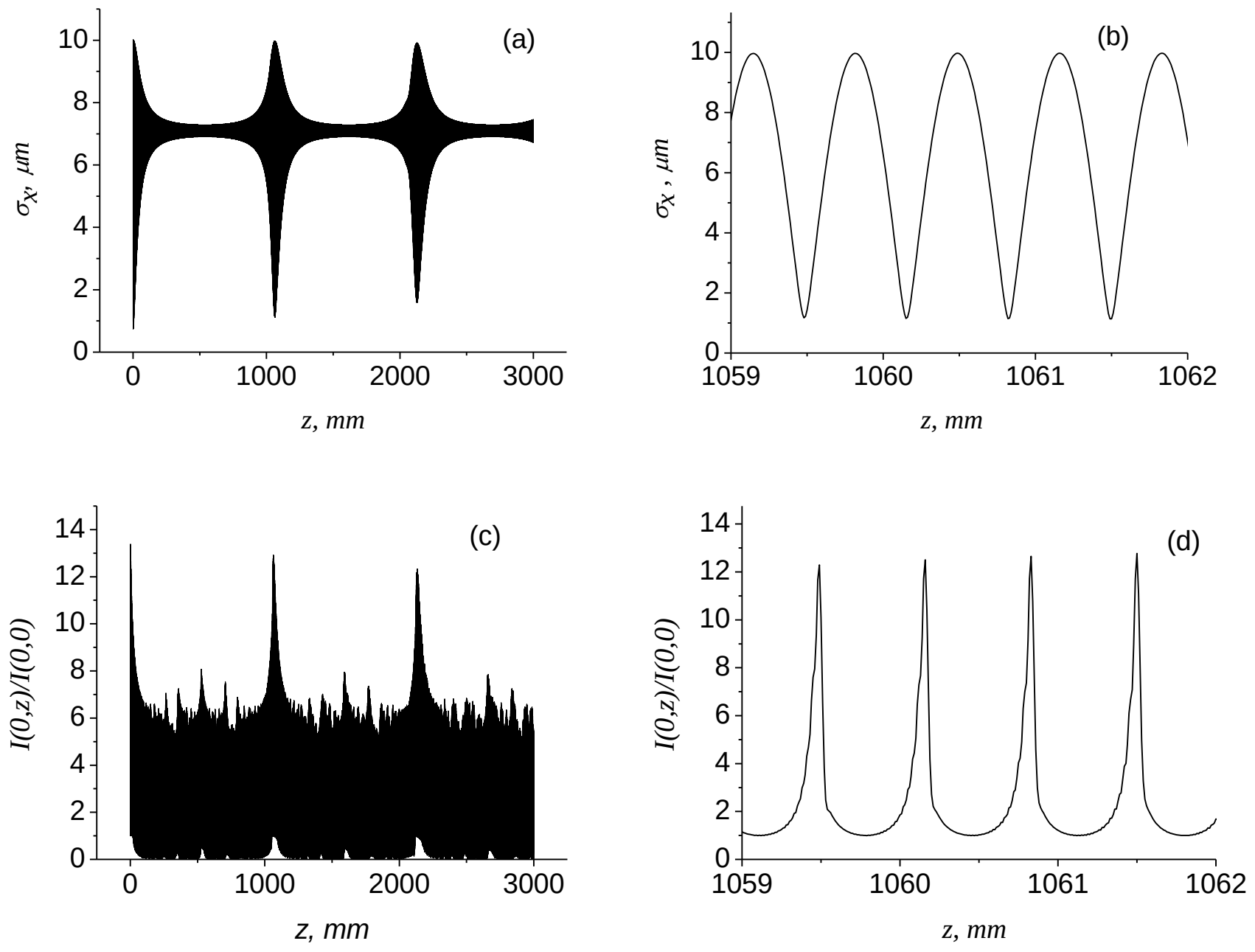

Fig. 2

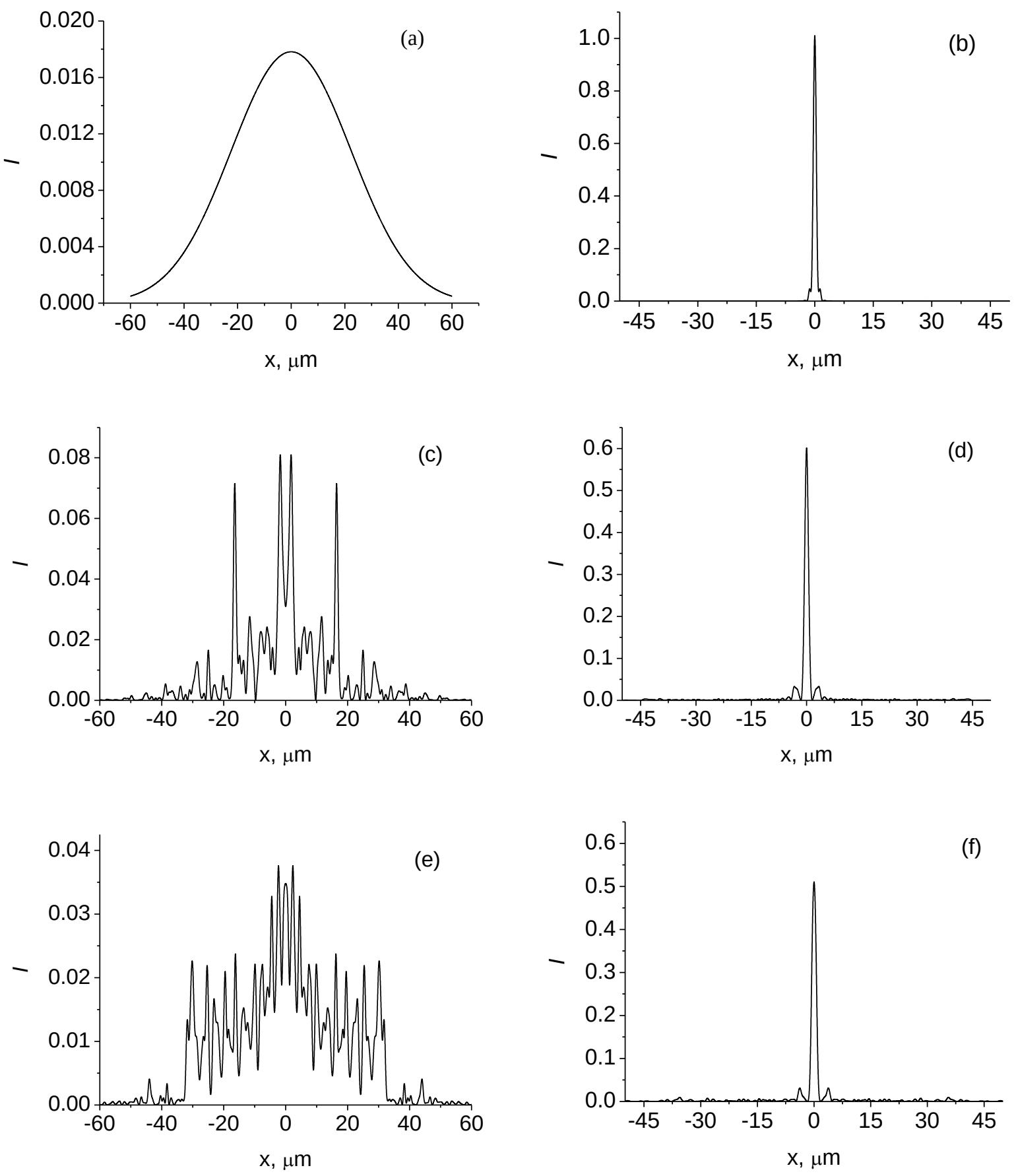

Fig. 3

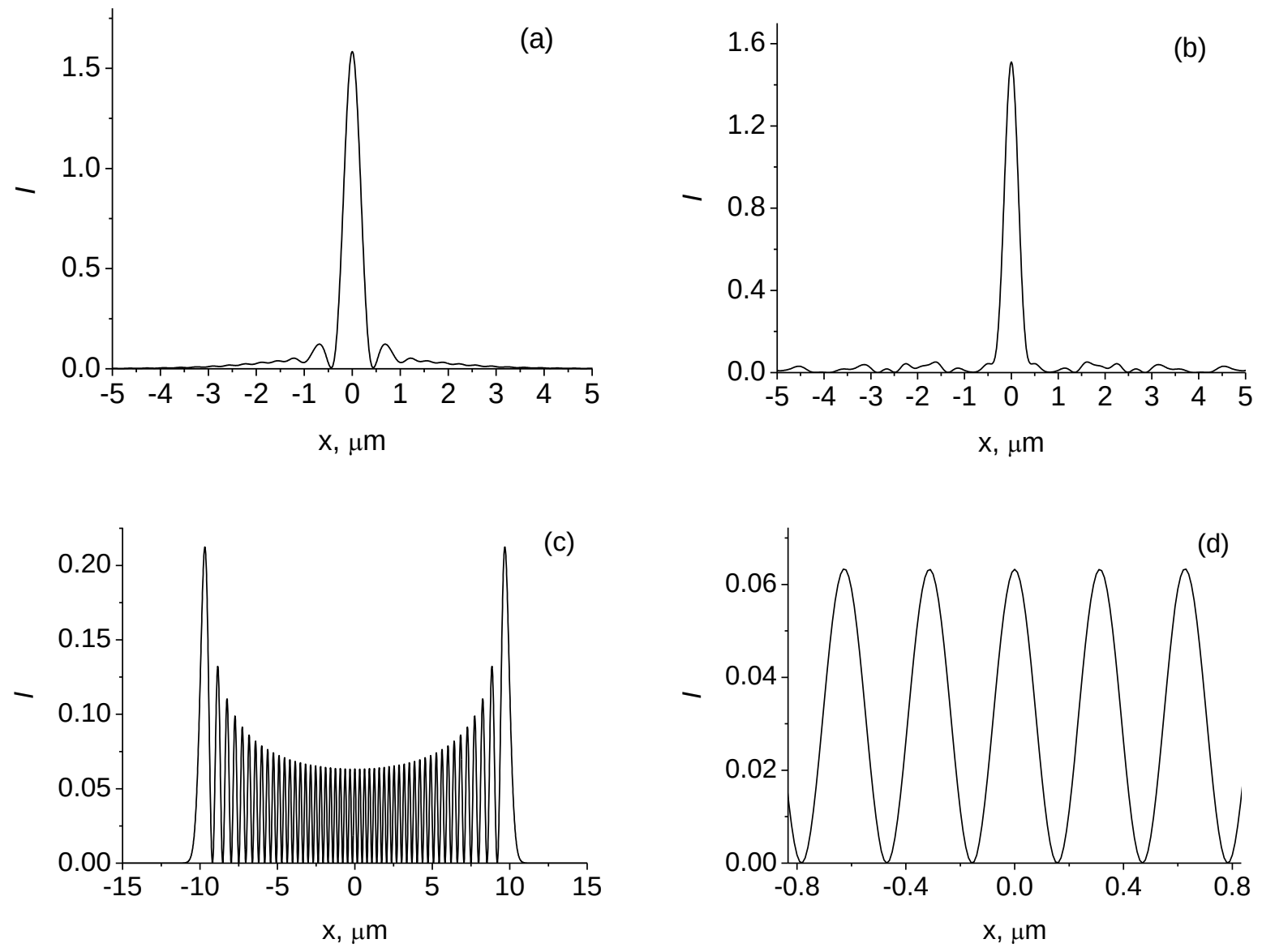

Fig. 4

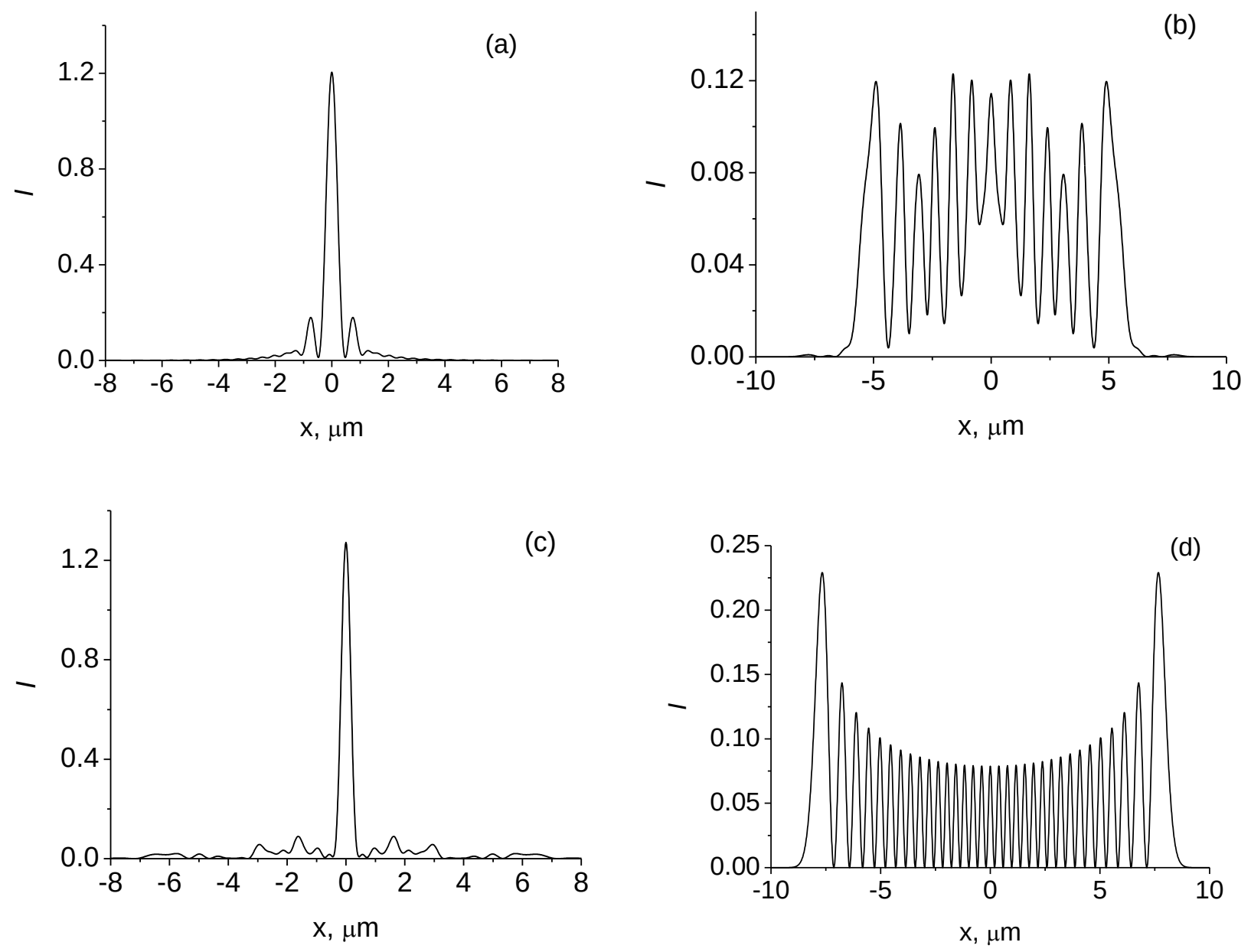

Fig. 5

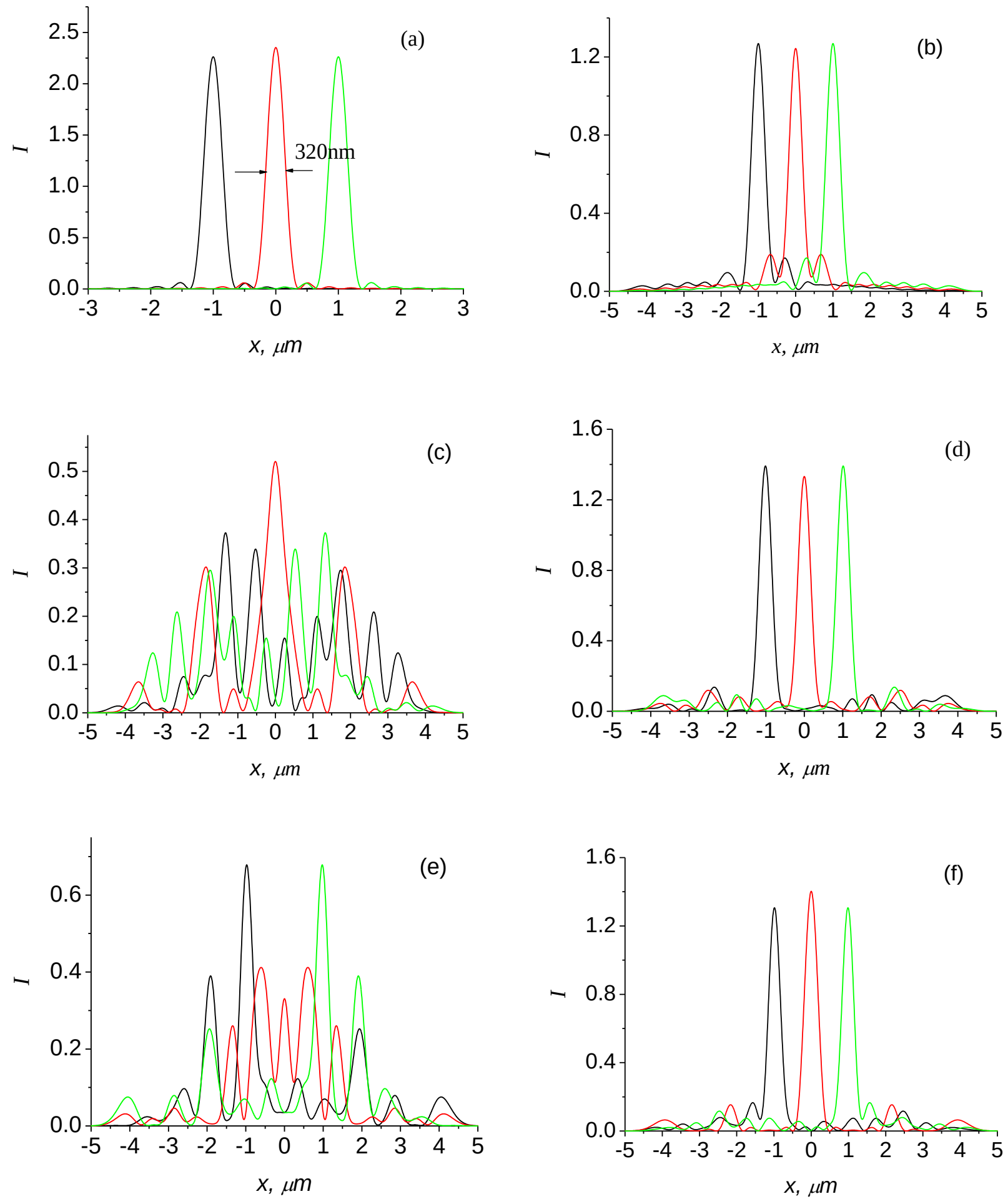